\documentclass{webofc}
\usepackage[varg]{txfonts}   
 
\begin{document}
%
\title{Hadron spectroscopy and interactions from lattice QCD}

\author{\firstname{Sasa} \lastname{Prelovsek}\inst{1,2}\fnsep\thanks{\email{ sasa.prelovsek@ijs.si}}}

\institute{Department of Physics, University of Ljubljana, Jadranska 19, Ljubljana, Slovenia  \\
           Jozef Stefan Institute, Jamova 39, 1000 Ljubljana, Slovenia            }

\abstract{
 Lattice QCD approach to study the hadronic resonances and exotic hadrons is described at an introductory level.  The main challenge is  that  these states decay strongly via one or more decay channels, and they often lie near thersholds. 
  Specific results for conventional and exotic hadrons are shown  to illustrate  the current status. 
    }
  
\maketitle

\section{Introduction} 

 Lattice QCD provides very accurate and reliable results for  hadrons that lie well below strong-decay threshold and  
 do not decay strongly. Over past few years, the lattice community provided also rigorous results for the resonances that can decay strongly via one-decay channel $R\to H_1H_2$. Interesting shallow bound states that lie just below $H_1H_2$ threshold were also investigated. Both  require the simulation of $H_1H_2$ scattering and extraction of the underlying scattering matrix.   A resonance that can strongly decay to  several channels presents an even more challenging problem,  
 which implies  for the majority of the exotic hadrons. First lattice results  were obtained very recently on this front as well. 
  Here I  illustrate the  lattice approach to this problem  in the introductory manner.    Only few selected results are shown to illustrate  various  types of  cases, while this is not a review  of available results.  
 
\section{Lattice methodology}

The physics information on a hadron (below, near or above threshold) is commonly extracted from the discrete energy spectrum in lattice QCD. 
The physical system for given quantum numbers is created from the vacuum $|\Omega\rangle$ using interpolator ${\cal O}_j^\dagger$  at time $t\!=\!0$ and the system propagates for time $t$ before being annihilated by ${\cal O}_i$.    To study a meson or an exotic-meson  with given $J^P$, one  can  use ${\cal O}\simeq \bar q	\Gamma q,~$ two-meson interpolators ${\cal O}=(\bar q \Gamma_1  q)(\bar q \Gamma_2  q),~$$(\bar q \Gamma_1  q)(\bar q \Gamma_2  q)~$ and  ${\cal O}=[\bar q \Gamma_1 \bar q][q\Gamma_2 q]$ with desired quantum numbers.  After the spectral decomposition the correlators are expressed in terms of the energies  $E_n$ of eigenstates $|n\rangle$ and their overlaps $Z_j^n$
\begin{equation}
\label{C}
C_{ij}(t)= \langle \Omega|{\cal O}_i (t) {\cal O}_j^\dagger (0)|\Omega \rangle=\sum_{n}Z_i^nZ_j^{n*}~e^{-E_n t}~,\qquad Z_i^n\equiv \langle \Omega|{\cal O}_i|n\rangle~.
\end{equation}
The correlators are evaluated on the lattice and their time-dependence allows to extract $E_n$ and $Z_n^i$  \cite{Michael:1985ne,Blossier:2009kd}. 

 The energy eigenstates $|n\rangle$ are   predominantly "one-meson" states (e.g. $\chi_{c1}(1P)$) or predominantly "two-meson" states (e.g. $D\bar D^*$) - in interacting theory they are mixtures of those.    Two-meson states  have a discrete spectrum due to the periodic boundary condition on the finite lattices.  If they do not  interact, then the momenta of each meson is $\vec{p}= \!\tfrac{2\pi}{L}\vec{N}$ with $\vec{N}\in {N}^3$, and the non-interacting energies of $M_1(\vec p)M_2(-\vec p)$    are $E^{n.i.}=E_1(p)+E_2(p)$ with $E_{1,2}(p)=(m_{1,2}^2+p^2)^{1/2}$.  The energies $E_n$ extracted from the lattice  are slightly shifted in presence of the interaction and the shift  provides rigorous information on the scattering matrix, as discussed in  Section \ref{sec:scat}. 
In experiment, two-meson  states correspond to the two-meson decay products with a continuous energy spectrum. 
  
   \section{Hadrons well below strong-decay threshold}
 
Masses are straightforwardly determined for the hadrons that can not decay strongly and lie well below strong decay threshold.
  These masses $m_i=E_i(P=0)$ are extracted  from the energies obtained with $\bar qq$ or $qqq$ interpolating fields,  which are extrapolated $a\to 0$, $V\to \infty$ and $m_q\to m_q^{phys}$.  The resulting masses for all such hadrons (i.e. $\pi$, $K$, $D$, $B$, $B^*$, $B_s$, $J/\psi$, $\eta_b$, $n$, $p$, $\Lambda$ ,...) are in very good agreement with the experimental measurements. The main remaining uncertainty is the omission of charm and beauty annihilation for heavy quarkonia.  
   
  \section{Scattering approach to resonances and bound states }\label{sec:scat}
  
 Most of the hadrons are resonances,  they are located above strong decay threshold(s) and decay strongly. 
   In the energy region near or above threshold, the masses of bound-states and resonances have to be inferred from the infinite-volume scattering matrix of the one-channel (elastic) or multiple-channel (inelastic) scattering.  The simplest example is a one-channel elastic scattering  in partial wave $l$, where   the scattering matrix  is parametrized in terms of the phase shift $\delta_l(p)$ and satisfies unitarity $SS^\dagger=1$
\begin{equation}
\label{T}
S(E)=e^{2i\delta_l(E)}\ , \quad S(E)=1+2iT(E)\ , \quad T(E)=\frac{1}{\cot(\delta_l(E))-i}~.
\end{equation}  
  L\"uscher has shown that the energy  $E$ of two-meson eigenstate in finite volume $L$ gives the elastic phase shift $\delta(E)$ at that energy in infinite volume  \cite{Luscher:1991cf}.  This relation and 
 its generalizations are at the core of extracting rigorous  information about the scattering from the recent lattice simulation.  It leads  $\delta(E)$ only for specific values of $E$ since  
  since spectrum  of two-meson eigenstates is discrete.  The $\delta(E)$ or $T(E)$ (\ref{T})  provide the masses of resonances and bound states:  
  \begin{itemize}
\item  In the vicinity of a {\it hadronic resonance} with mass $m_R$ and width $\Gamma$, the  cross section $\sigma \propto |T(p)|^2$ has a Breit-Wigner-type shape with  $\delta(s=m_R^2)=\tfrac{\pi}{2}$
\begin{equation}
\label{R}
T(p)=\frac{-\sqrt{s}~ \Gamma(p)}{s-m_R^2+i \sqrt{s}\, \Gamma(p)}=\frac{1}{\cot\delta(p)-i}\ , \quad
\Gamma(p)=g^2\,\frac{p^{2l+1}}{s} . 
\end{equation}  
The fit of $\delta_l(p)$  renders $m_R$ and $g$ or $\Gamma$. It is customary to compare $g$ rather than $\Gamma$ to experiment, since $\Gamma$ depends on the phase space.  L\"usher's  approach has been verified on several conventional mesonic resonances. 
\item The {\it bound state (B)} in $M_1 M_2$ scattering is realized when $T(p)$ has a pole at $p_B^2<0$ or  $p_B= i|p_B|$
\begin{equation}
\label{B}
T=  \frac{1}{\cot(\delta_l(p_B))-i}=\infty\ , \ \  \cot(\delta(p_B))=i\ ,\quad m_B=E_{H_1}(p_B)+E_{H_2}(p_B)~.  
\end{equation}
 The  location of an s-wave shallow bound state can be obtained by parametrizing $\delta_0$ near threshold and finding $p_B$ which satisfies $\cot(\delta(p_B))=i$.   
 \end{itemize}
 
  \begin{figure}[htb] 
\begin{center}
\includegraphics[width=0.35\textwidth,clip]{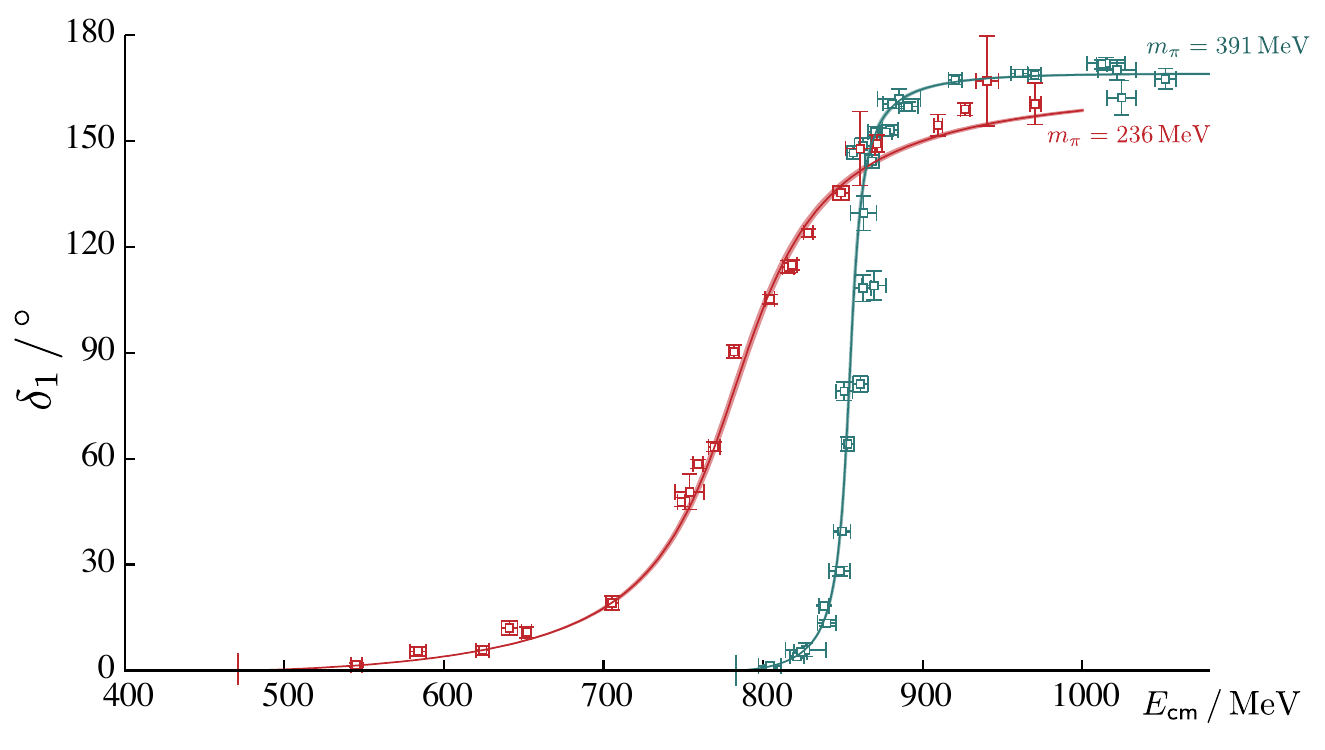} $\quad$
\includegraphics[width=0.6\textwidth,clip]{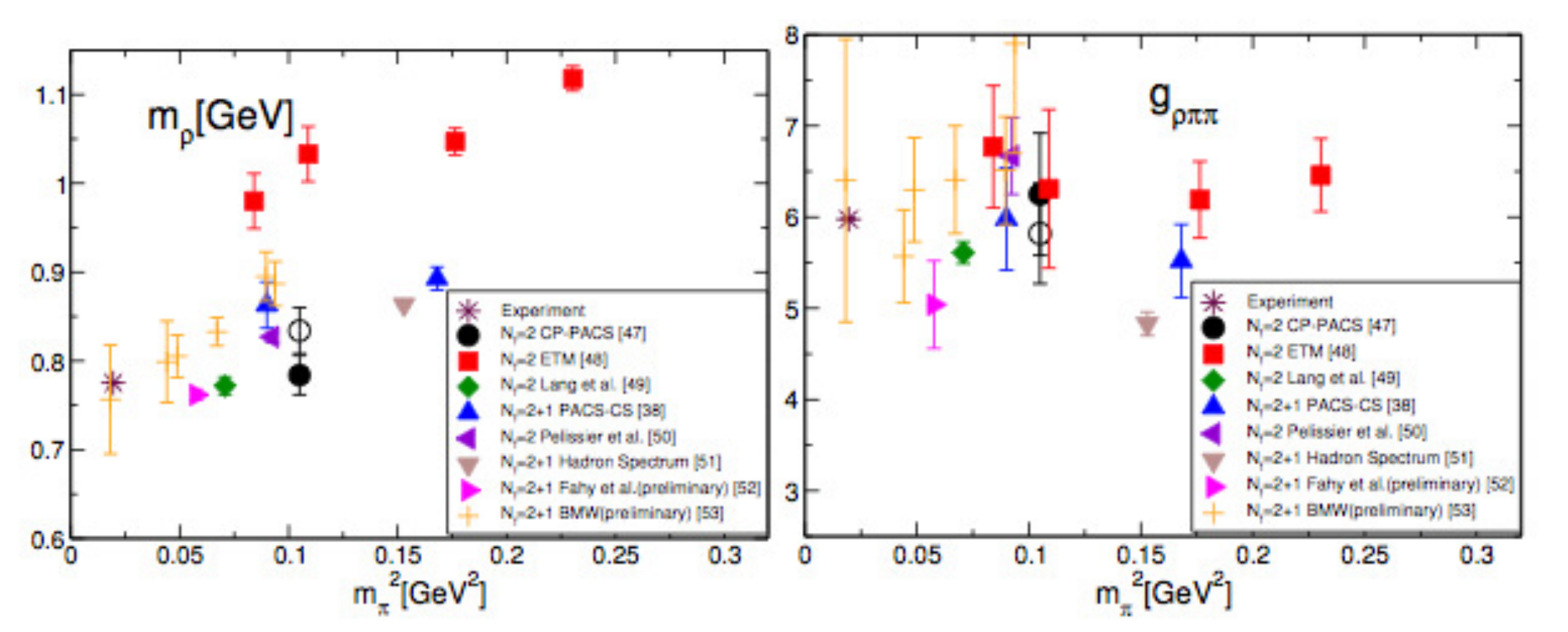}  
\caption{\label{fig:rho}    Left: The $\pi\pi$ phase shift in the $\rho$-meson channel     from the simulation  by the Hadron Spectrum Collaboration \cite{Dudek:2012xn,Wilson:2015dqa}.    Right: The compilation  \cite{Yamazaki:2015nka} of the   masses and widths for the $\rho$ meson from recent lattice simulations, where the $g$ parametrizes the width as $\Gamma=g^2 p^3/(6\pi s)$.   }
\end{center}
\end{figure}

    \begin{figure}[htb] 
\begin{center}
\includegraphics[width=0.4\textwidth,clip]{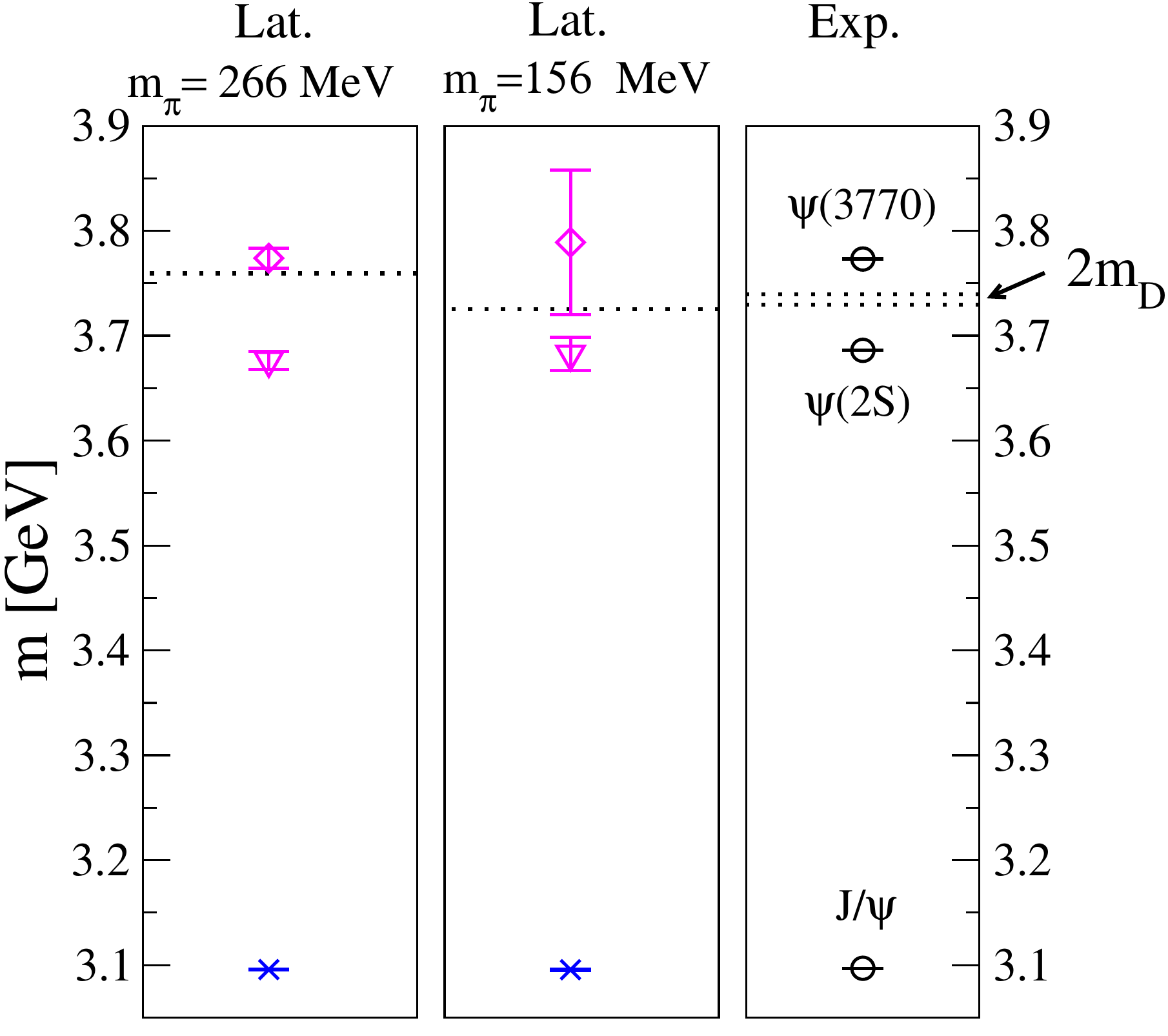} $\quad$ 
\includegraphics[width=0.4\textwidth,clip]{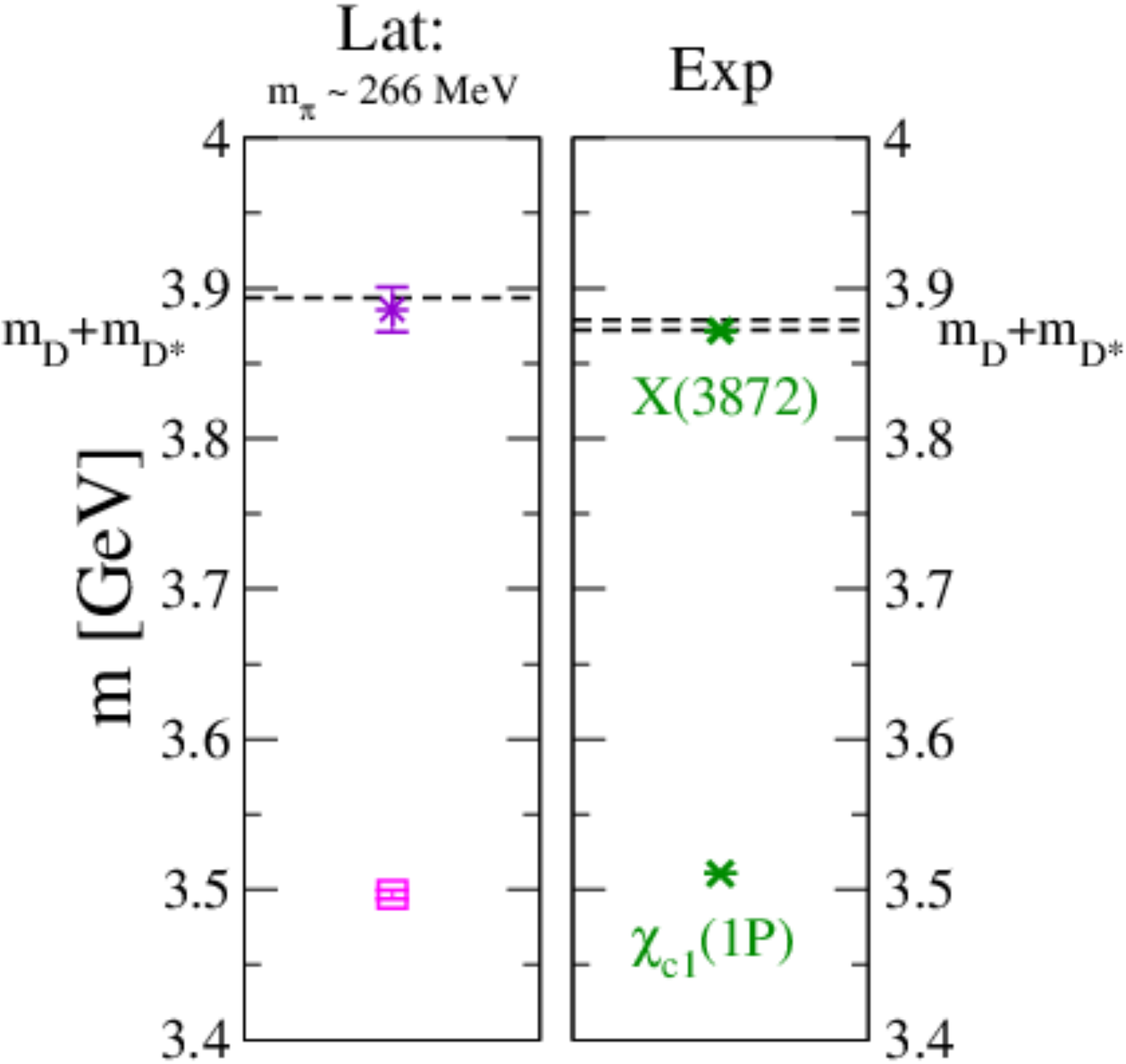}  
\caption{\label{fig:vector_charmonia}   Left: The  lattice spectrum of the vector charmonia from \cite{Lang:2015sba}: the  diamond  denotes the resonance mass of  $\psi(3770)$, while the triangle denotes the pole mass of the bound state $\psi(2S)$; both are obtained from $D\bar D$ scattering matrix.   Right: The location of $X(3872)$ with $I=0$ which emerges as shallow bound state in a simulation of  $D\bar D^*$ scattering   that includes also diquark antidiqark interpolating fields \cite{Padmanath:2015era}.   }
\end{center}
\end{figure}

 \begin{figure}[htb] 
\begin{center}
\includegraphics[width=0.38\textwidth,clip]{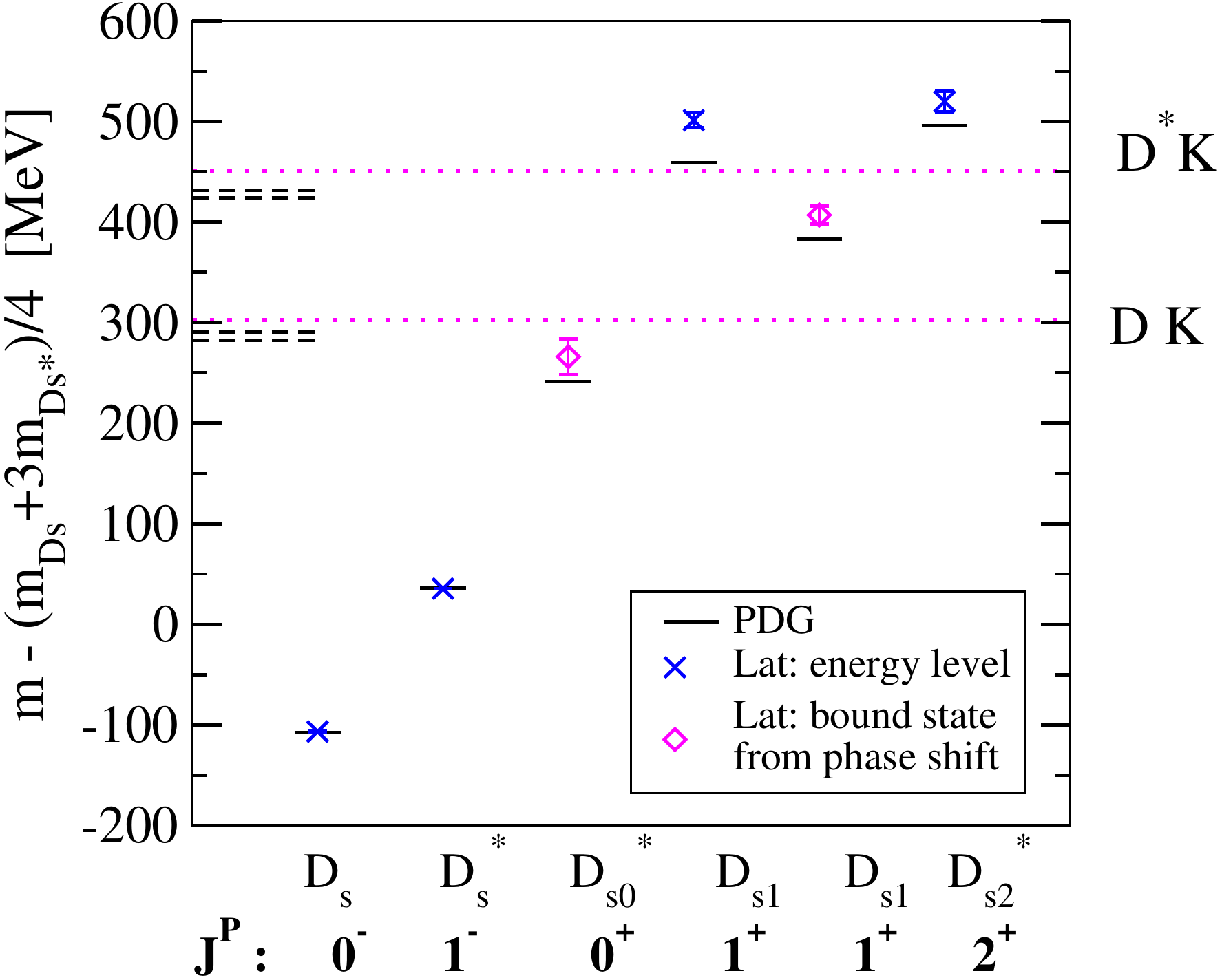} $\quad$
\includegraphics[width=0.35\textwidth,clip]{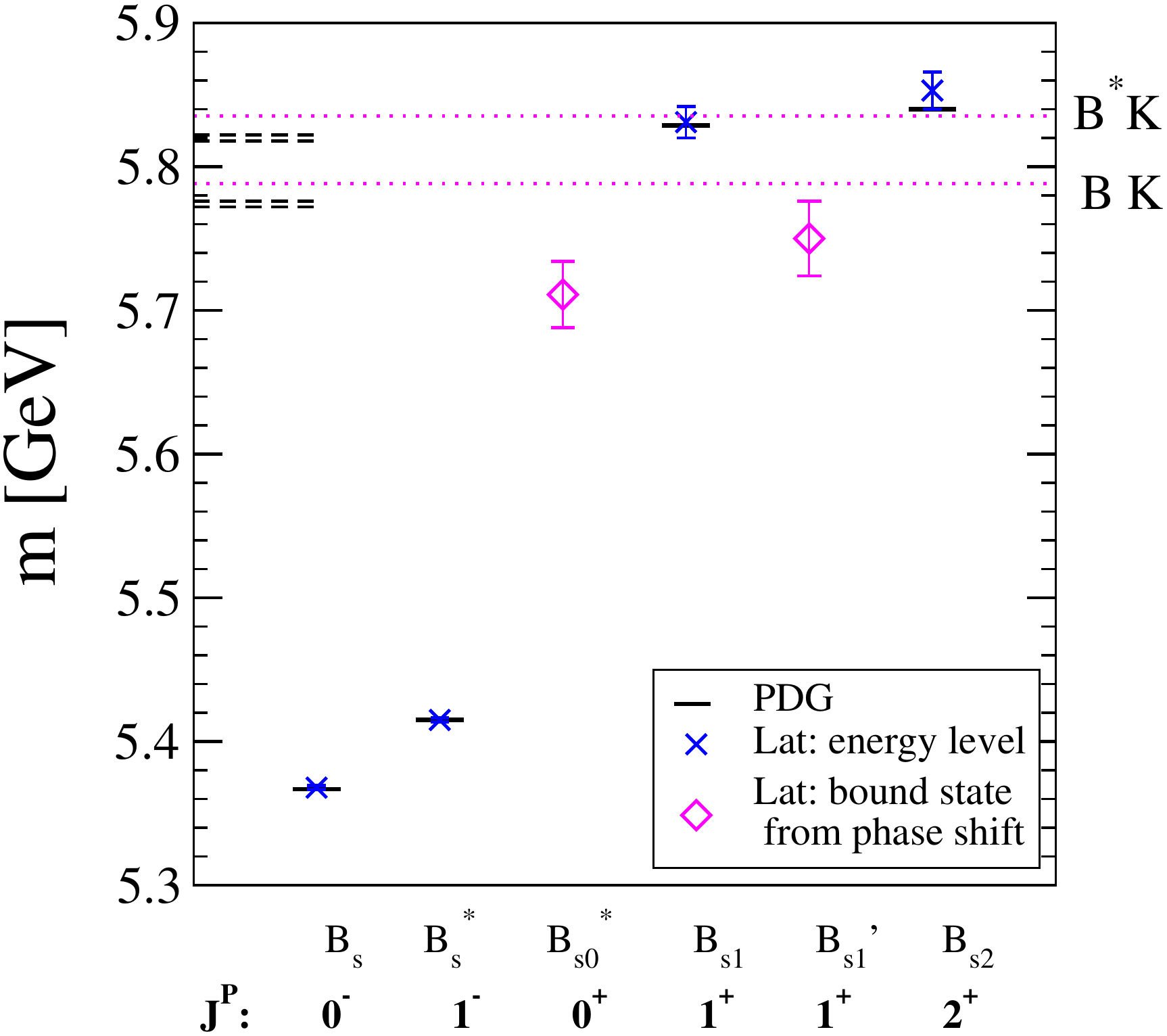}  
\caption{\label{fig:Ds_Bs}     The spectra of $D_s$ \cite{Mohler:2013rwa,Lang:2014yfa} and $B_s$ \cite{Lang:2015hza} mesons resulting from $N_f=2+1$ lattice simulation on PACS-CS ensembles with $m_\pi\simeq 156~$MeV. The black horizontal lines are experimental masses.   }
\end{center}
\end{figure}

  \section{ Selected resonances from one-channel scattering } 
  
  \underline{$\rho$ meson}: Only one resonance - namely the $\rho$ meson - has been  rigorously studied by a number of lattice collaborations, reviwed in  \cite{Yamazaki:2015nka}. Figure \ref{fig:rho} shows an example of  extracted $\pi\pi$ phase shift for $L=1$ and $I=1$, obtained by the Hadron Spectrum Collaboration  \cite{Dudek:2012xn,Wilson:2015dqa} for two pion masses. The Breit-Wigner-type fit renders  $\rho$ resonance mass and width \cite{Dudek:2012xn}    shown by the brown triangle in Fig. \ref{fig:rho}, where other lattice results are also shown. \\
  
    \underline{Charmonium resonances}:  The charmonium spectrum above open charm threshold  contains a number of exotic candidates.  It is  important to verify the approach   on  the conventional   states  before    addressing   the  exotic states.  Until recently, all charmonia above open-charm threshold were treated ignoring the strong decay 
  to a pair of charmed mesons. 
  The first exploratory  simulation aimed at determining the masses as well as the decay widths  of these resonances was presented in \cite{Lang:2015sba}.  The Breit-Wigner-type fit of the $D\bar D$ phase shift in $p$-wave leads to the resonance   $\psi(3770)$ mass in agreement with experiment, as shown in Fig.  \ref{fig:vector_charmonia}. The extracted  $\psi(3770)\to D\bar D$ decay width $\Gamma=g^2 p/s$ with  $g_{m_\pi=266\;\mathrm{MeV}}^{lat}=19.7\pm 1.4$    is also close to  $g^{exp}=18.7\pm 1.4$.    The $\psi(2S)$ in Fig.    \ref{fig:vector_charmonia} \label{fig:vector_charmonia} appears as a bound state pole below threshold.  In the scalar channel, only the ground state $\chi_{c0}(1P)$ is understood and there is no commonly accepted candidate  for its first excitation  $\chi_{c0}(2P)$. The lattice results on the $D\bar D$ scattering in s-wave  \cite{Lang:2015sba} leave open puzzles and call for further work on this problem.

   \section{ Selected shallow bound states from one-channel scattering } 

\underline{Scalar and axial mesons $D_s$ and $B_s$}:  The quark models expected $D_{s0}^*(2317)$ and $D_{s1}(2460)$  above  $DK$ and $D^*K$ thresholds, but they were experimentally found slightly below them.  These are one of few shallow  bound states in the meson sector, and therefore deserve special attention.  The first lattice simulation that takes the effect of these thresholds into account   was presented in \cite{Mohler:2013rwa,Lang:2014yfa}. It employs $DK$ and  $D^*K$ interpolating fields in addition to the $\bar sc$.   The $D^{(*)}K$ phase shift is extracted   and then interpolated  in the region close to threshold. 
A pole of the scattering matrix  is found below the thresholds where the relation (\ref{B}) is satisfied. The poles are  related to   $D_{s0}^*(2317)$  and $D_{s1}(2460) $ bound states and are close to the masses of  experimentally observed states, as shown    in Fig. \ref{fig:Ds_Bs}.  The scalar and axial $B_s$ mesons have not been discovered in experiment yet, and in this case we made a prediction for their masses using analogous simulation of $B^{(*)}K$ scattering.  We find poles related to $B_{s0}$ and $B_{s1}$ mesons below $BK$ and $B^*K$ thresholds \cite{Lang:2015hza},   as shown in Fig. \ref{fig:Ds_Bs}. \\

\underline{$X(3872)$}: The $X(3872)$ lies  experimentally on $D^0\bar D^{0*}$ threshold and its existence on the lattice can not be established without taking into account the effect of this threshold. This was first done by simulating $D\bar D^*$ scattering  in \cite{Prelovsek:2013cra}, where   a pole in $D\bar D^*$ scattering matrix was found just below the threshold in $I(J^{PC})=0(1^{++})$ channel  (\ref{B}). The pole was associated with $X(3872)$ \cite{Prelovsek:2013cra} and was also confirmed by a simulation with HISQ action \cite{Lee:2014uta}.  
The more recent simulation  \cite{Padmanath:2015era}  included also the  diquark anti-diquark  interpolating fields  and the location of the $X(3872)$ pole is shown in Fig. \ref{fig:vector_charmonia}.  
  
The lattice study \cite{Padmanath:2015era} investigated which Fock components are essential for appearance of $X(3872)$ with $I=0$ on the lattice. The energy eigenstate  related to $X(3872)$ appears in the simulation only if  $D\bar D^*$ as well as $\bar cc$ interpolating fields are employed.  The $X(3872)$ does not appear in absence of  $\bar cc$ interpolators, even if  (localized)  interpolators $[\bar c\bar q]_{3_c}[cq]_{\bar 3_c}$ or $[\bar c\bar q]_{6_c}[cq]_{\bar 6_c}$  are in the interpolator basis. This indicates that $\bar cc$ Fock component is most likely more essential for $X(3872)$ than the diquark-antidiquark one. \\

\underline{Pentaquark bound state}: The NPLQCD collaboration finds an interesting evidence for a $\eta_c N$ bound state approximately $20~$MeV below $\eta_c N$ threshold   \cite{Beane:2014sda} ($N$ denotes nucleon). To my knowledge, this is the only pentaquark candidate containing $\bar cc$ from lattice studies up to now. As the simulation is done at $SU(3)$ flavour symmetric point corresponding to $m_\pi\simeq 800~$MeV, it is not clear yet whether this bound state persists to physical $m_\pi$.  LHCb has recently found two pentaquark  resonances   in $J/\psi p$, located about $400~$MeV above threshold \cite{Aaij:2015tga}. The lattice simulation of those is much more challenging due to several open channels and has not been performed yet. 

 \begin{figure}[htb] 
\begin{center}
\includegraphics[width=0.45\textwidth,clip]{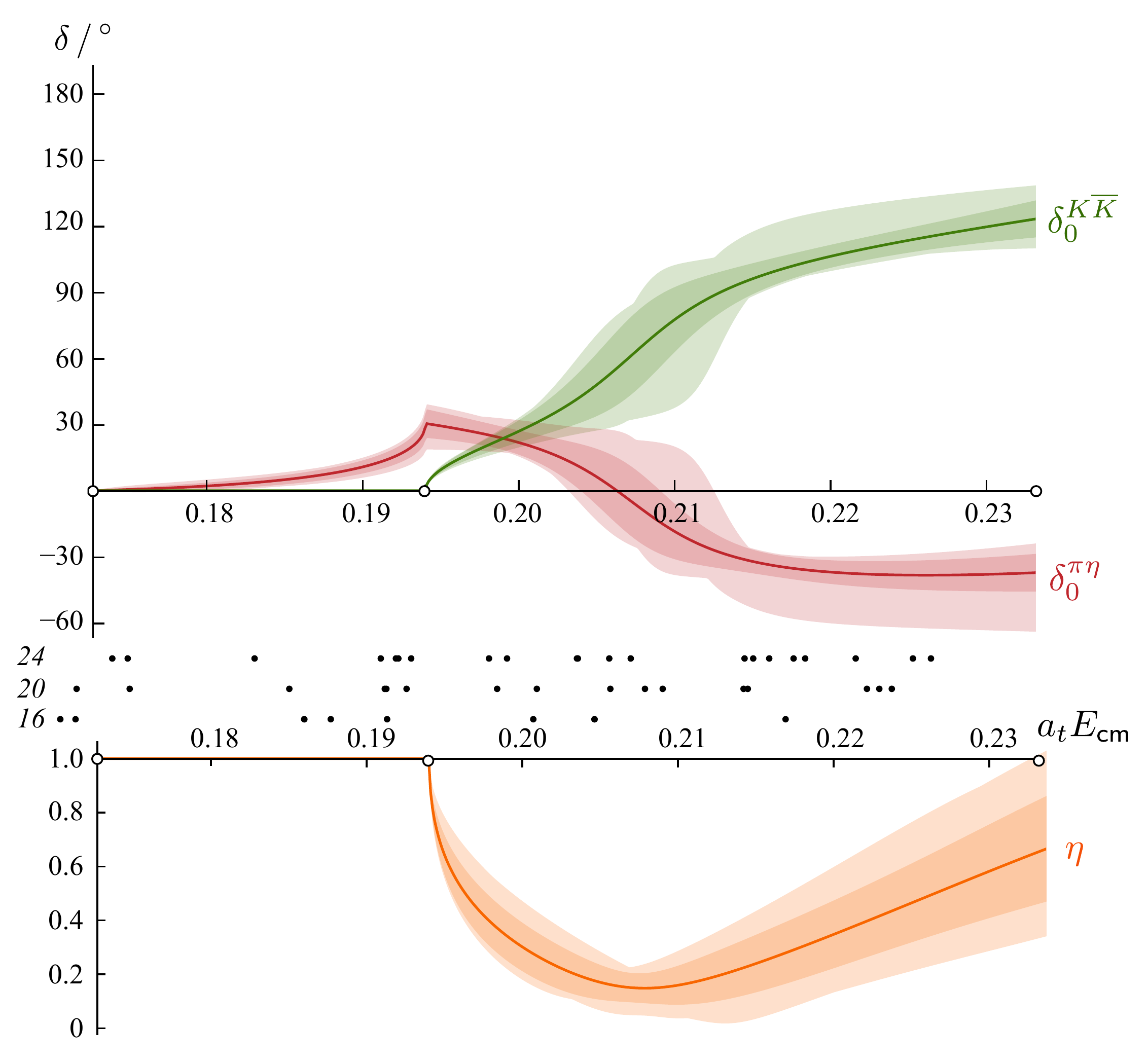}  
\caption{\label{fig:a0}      The phases and the inelasticity of the $ K\bar K - \pi \eta$ coupled channel scattering in the channel where $a_0$ resonance appears in experiment.  The results are obtained from a $N_f=2+1$ lattice simulation at $m_\pi\simeq 400~ $MeV by the Hadron Spectrum Collaboration \cite{Dudek:2016cru}. }
\end{center}
\end{figure}

 \begin{figure}[htb] 
\begin{center}
\includegraphics[width=0.75\textwidth,clip]{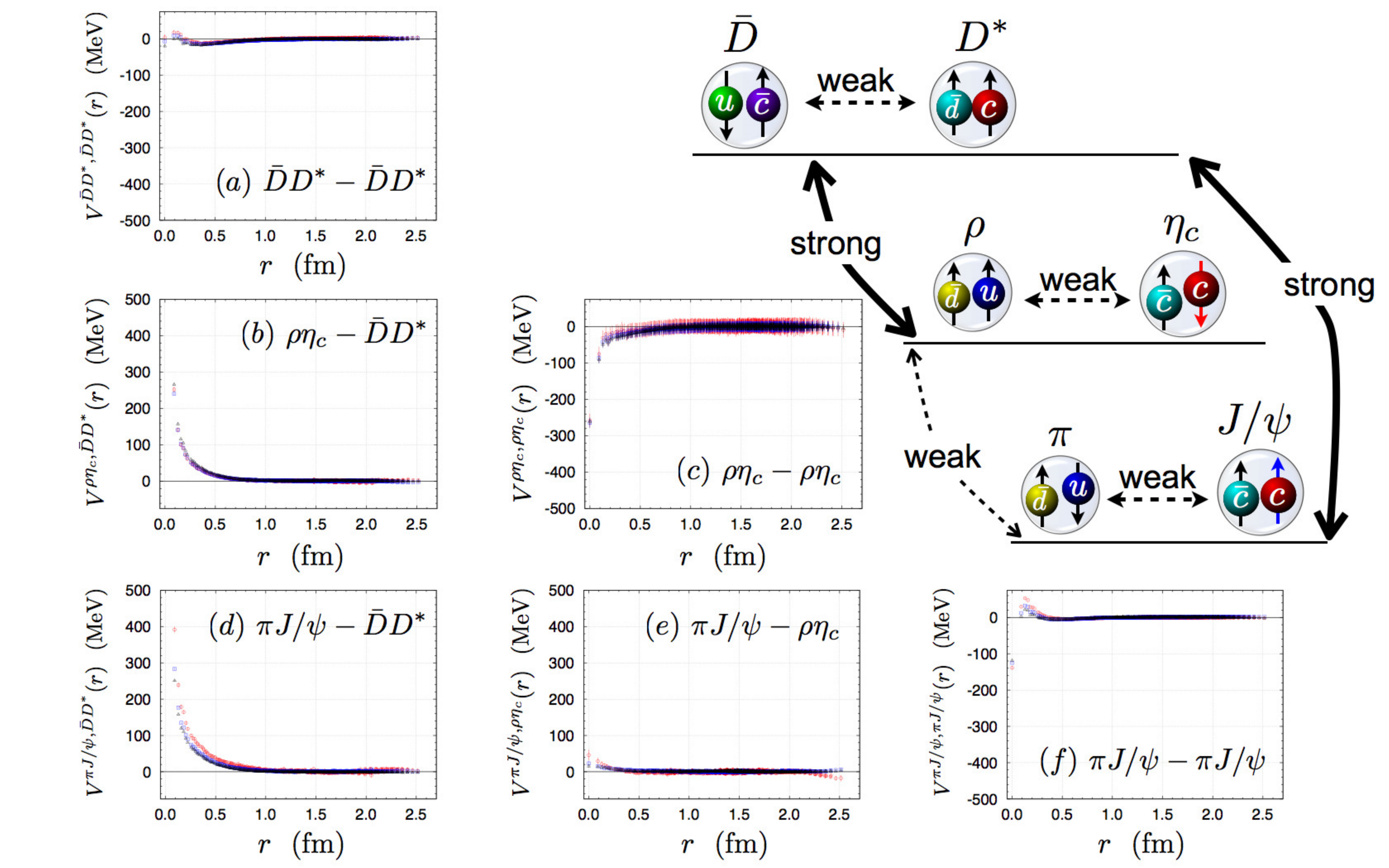} 

$\quad$
 
\includegraphics[width=0.32\textwidth,clip]{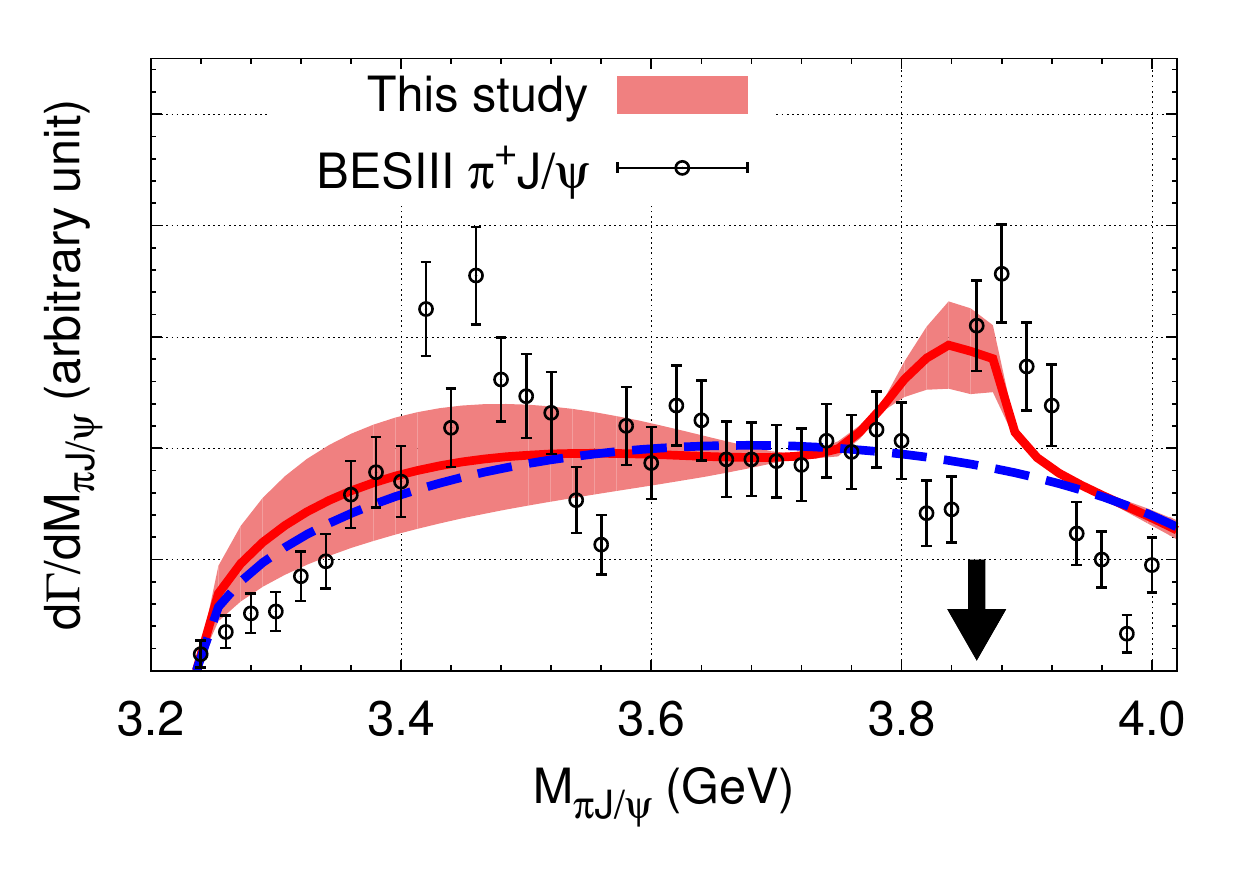} 
\includegraphics[width=0.32\textwidth,clip]{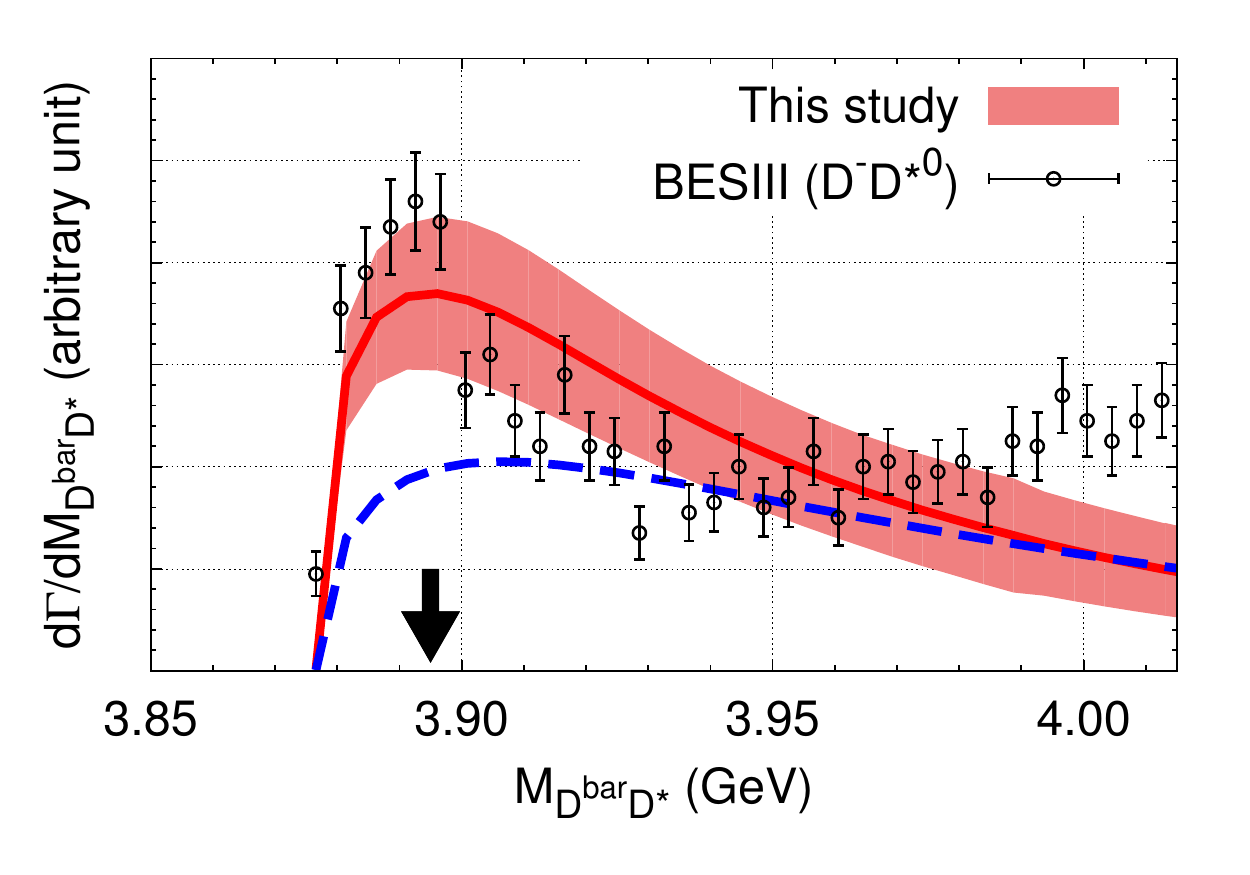}  
\includegraphics[width=0.32\textwidth,clip]{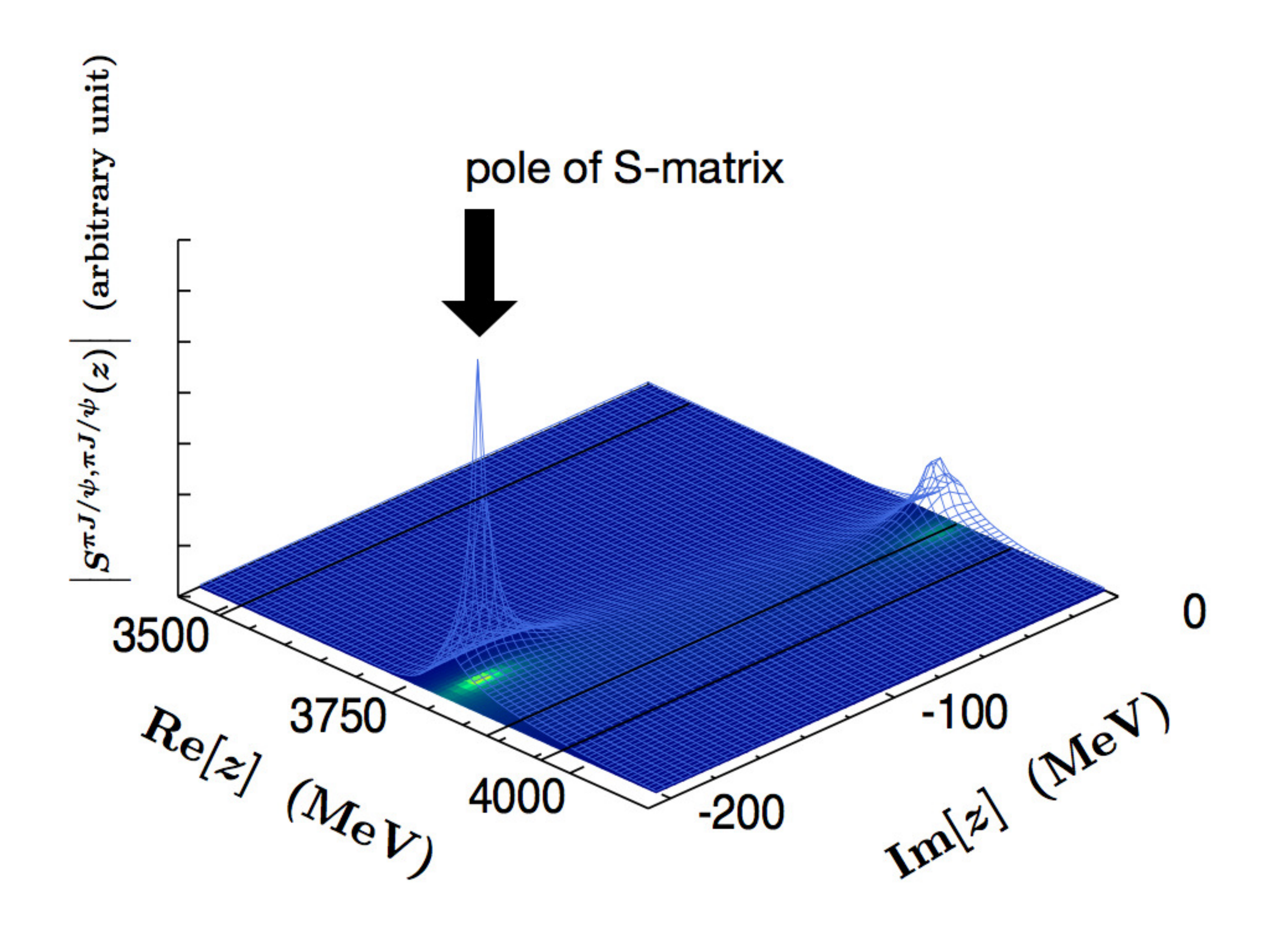}  
\caption{\label{fig:Zc_halqcd}  The results related to $Z_c^+(3900)$ from a lattice simulation \cite{Ikeda:2016zwx}, based on the  HALQCD method.    Top: potentials $V(r)$ for three channels. Bottom left: red lines present the lattice QCD results for  $Y(4260) \to J/\psi \pi\pi$ and  $Y(4260) \to D \bar D^*\pi$ invariant masses, while blue lines present results if $J/\psi\pi$-$D\bar D^*$ coupling is turned off.     Bottom right: Poles of the scattering matrix in the complex energy plane.   }
\end{center}
\end{figure}

 \section{  Towards the coupled-channel scattering  and exotic hadrons}  
 
 Most of the   exotic candidates are experimentally found  above several thresholds, and can decay in more than one hadronic final state. This problem  requires the determination of the scattering matrix for the coupled-channel scattering. This is more challenging for a rigorous treatment than a simulation of one-channel (elastic) scattering. \\
 
 \underline{ $a_0$ resonance:} The coupled-channel scattering matrix was extracted  from the lattice simulation for the first time only recently \cite{Dudek:2014qha}, where the generalized L\"uscher's method was used. Figure \ref{fig:a0} shows the results  for the scattering phases and inelasticity that parametrize the $2\times 2$ scattering matrix in the $a_0$ channel   \cite{Dudek:2016cru}. The analytic  proposals to parametrize the coupled-channel scattering matrix in order to enable its lattice determination have been presented earlier, for example in \cite{Lage:2009zv,Bernard:2010fp}.   \\
 
  \underline{$Z_c^+(3900)$}:  The lattice search for the manifestly exotic states $Z_c^+$ with flavour content $\bar cc\bar du$ and $I^G(J^{PC})=1^+(1^{+-})$ is very challenging since 
the experimental candidate can strongly decay to several final states  $D\bar D^*$, $J/\psi \pi$ and  $\eta_c \rho$.  The corresponding $3\times 3$ scattering matrix has never been determined using the reliable L\"uscher's method. 

The HALQCD collaboration used the HALQCD approach to extracted the coupled-channel scattering matrix, which is not commonly accepted as an exact treatment, and has not been verified on the conventional resonances.   First the potential $V_{\pi J/\psi -\pi J/\psi}(r)$ related to Nambu-Bethe-Salpeter equation is determined between the $J/\psi$ and $\pi$  as a function of their separation $r$ \cite{Ikeda:2016zwx}. This is presented in   Fig. \ref{fig:Zc_halqcd} together 
  with potential for the other two channels, as well as potential corresponding to coupling between different channels according to coupled-channel HALQCD formalism \cite{Aoki:2012bb}.  The off-diagonal potential between channels $\pi J/\psi$ and $D\bar D^*$ is larger than other potentials, which   seems to indicate a sizable coupled channel effect near $D\bar D^*$ threshold. The  potentials render $3\times 3$ scattering matrix for three coupled channels \cite{Ikeda:2016zwx}. This is then used to determine the three body decay 
  $Y(4260) \to J/\psi \pi\pi$ and  $Y(4260) \to D \bar D^*\pi$ in a semi-phenomenological way. 
  The result in Fig. \ref{fig:Zc_halqcd} indeed shows a peak around $Z_c$ mass.  If the coupling between $J/\psi \pi$ and $D\bar D^*$ is turned off, the  peak disappears as shown by  blue dashed line.  It indicates that the coupling of $J/\psi \pi$ and $D\bar D^*$  channels seems to be crucial for the existence of $Z_c$. This needs to be verified also  by the L\"uscher's method.  
  
   The first step towards the treatment of this problem with the L\"uscher method is to determine the energies of eigenstates for the $Z_c$ system. This has been done in \cite{Prelovsek:2014swa,Lee:2014uta}. The number of lattice eigenstates in the relevant energy region was found to be equal to the number of expected two-hadron states in the absence of interactions, and no additional eigenstate was found.  This suggests that $Z_c$ is perhaps not a  resonance in the usual sense, i.e. it is not associated with  a pole above $D\bar D^*$ threshold near the real axis. Indeed, the HALQCD collaboration  has  not found such a pole, as shown in Fig. \ref{fig:Zc_halqcd}.  More analytical work and lattice simulations are required to conclude on  this problem. \\
   
\underline{$X(4140)$}:  The experimental candidate $X(4140)$ with hidden strangeness was confirmed in $J/\psi \phi$ invariant mass  by LHCb recently \cite{Aaij:2016iza,Aaij:2016nsc}. The amplitude analysis favors quantum numbers $J^{PC}=1^{++}$ for it.  The rigorous treatment would require extraction of the scattering matrix for at leat two channels $J/\psi \phi$ and $D_s\bar D_s^*$.  
The simplified lattice search was performed in this channel with $D_s\bar D_s^*$, $J/\psi \phi$, $[\bar c\bar s][cs]$  interpolators. The number of lattice eigenstates in the relevant energy region was found to be equal to the number of expected two-hadron states in the absence of interactions. No additional eigenstate was found near $4.1~$GeV, which suggests   that $X(4140)$ is perhaps not a resonance with a pole  above  $J/\psi \phi$ threshold near the real axis. The experimental study  \cite{Aaij:2016nsc}  found that the amplitudes  could be fitted within the  cusp scenario \cite{Swanson:2015bsa}, which might be consistent with 
 absence of a resonance pole  near $E_{cm}\simeq m_{X}-i \Gamma_X/2$.    
  
\section{Summary}

   During the past few years, the lattice QCD   provided rigorous results on the hadrons that lie close and above strong-decay thresholds. 
   Verification of the approach on the conventional hadrons is the first important step towards understanding exotic ones.  
  Most of the experimental exotic candidates can strongly decay via several  decay channels and lie near thresholds. This presents a challenge for rigorous treatment from first principles, but important encouraging steps in this direction have been accomplished. 
     
  \vspace{0.5cm}
  
  {\bf Acknowledgments }
  
  I would like to thank   C.B. Lang, D. Mohler, L. Leskovec, M. Padmanath and R. Woloshyn for pleasant collaboration on these physics problems.  Support from Slovenian Research Agency ARRS under project N1-0020 is acknowledged.




\end{document}